\documentclass[twocolumn]{aastex62}
\usepackage{color}
\usepackage[titletoc]{appendix}
\usepackage{amsmath}
\usepackage{amssymb}
\usepackage{mathtools}
\usepackage{upgreek}
\usepackage{float}
\usepackage{comment}
\usepackage{enumitem}
\usepackage{natbib}
\usepackage{graphicx}
\usepackage{bm}
\usepackage{totcount}
\usepackage{multirow}
\usepackage{pifont}

\newtotcounter{citnum} 
\def\oldbibitem{} \let\oldbibitem=\bibitem
\def\bibitem{\stepcounter{citnum}\oldbibitem}

\shortauthors{Sethi \& Millholland}
\shorttitle{Linking Orbital Misalignments \& Tidal Inflation}
\begin{document} 
\title{Tidal Inflation is Stronger for Misaligned Neptune-Sized Planets Than Aligned Ones}
\author[0000-0002-6576-3346]{Ritika Sethi}
\affiliation{Department of Physics, Massachusetts Institute of Technology, Cambridge, MA 02139, USA}
\affiliation{MIT Kavli Institute for Astrophysics and Space Research, Massachusetts Institute of Technology, Cambridge, MA 02139, USA}
\email{rsethi@mit.edu}
\author[0000-0003-3130-2282]{Sarah C. Millholland}
\affiliation{Department of Physics, Massachusetts Institute of Technology, Cambridge, MA 02139, USA}
\affiliation{MIT Kavli Institute for Astrophysics and Space Research, Massachusetts Institute of Technology, Cambridge, MA 02139, USA}
\email{sarah.millholland@mit.edu}

\begin{abstract}
Recent observations have revealed an intriguing abundance of polar-orbiting Neptune-sized planets, many of which exhibit unusually inflated radii. While such misaligned orbits point to a complex dynamical history, the connection between their orbital orientations and planetary structures remains poorly understood. In this study, we analyze a sample of 12 misaligned and 12 aligned planets using structure models that incorporate tidal heating. We use various statistical tests to demonstrate -- with at least $90\%$ confidence -- that misaligned planets experience more tidally-induced radius inflation compared to aligned planets. This inflation likely stems from their dynamically active histories, which often place them in close-in, eccentric, and highly inclined orbits. We further present a case study of WASP-107~b, an exceptionally inflated polar Neptune, and model its history using a simple coupled orbital and radius evolution approach. Our results place constraints on the planet's tidal quality factor that agree with recent JWST observations. 
\end{abstract}
\section{Introduction}
\label{sec: Introduction} 
It was long expected that planetary orbits should be well-aligned with their host star's equatorial planes due to their formation in thin disks. Surprisingly, observations have revealed that many exoplanets exhibit substantial \textit{spin-orbit misalignments} or \textit{stellar obliquities} ($\Psi$, the angle between a star's spin axis and a planet's orbital angular momentum vector) \citep{2005ApJ...631.1215W, 2015ARA&A..53..409W, 2022PASP..134h2001A}. These misalignments correlate with both stellar and planetary properties, suggesting a complex evolutionary history. Proposed mechanisms for misalignment include planet-planet scattering \citep{2002Icar..156..570M, 2008ApJ...686..580C, 2012ApJ...751..119B}, von Zeipel-Kozai-Lidov cycles \citep{2007ApJ...669.1298F, 2011Natur.473..187N}, secular resonances triggered by disk dissipation or stellar spin-down \citep{2018MNRAS.480.1402A, 2020ApJ...902L...5P}, and secular chaos \citep{2017MNRAS.470.1657H, 2019MNRAS.486.2265T}.

\begin{figure*}
    \centering
    \includegraphics[width=0.99\linewidth]{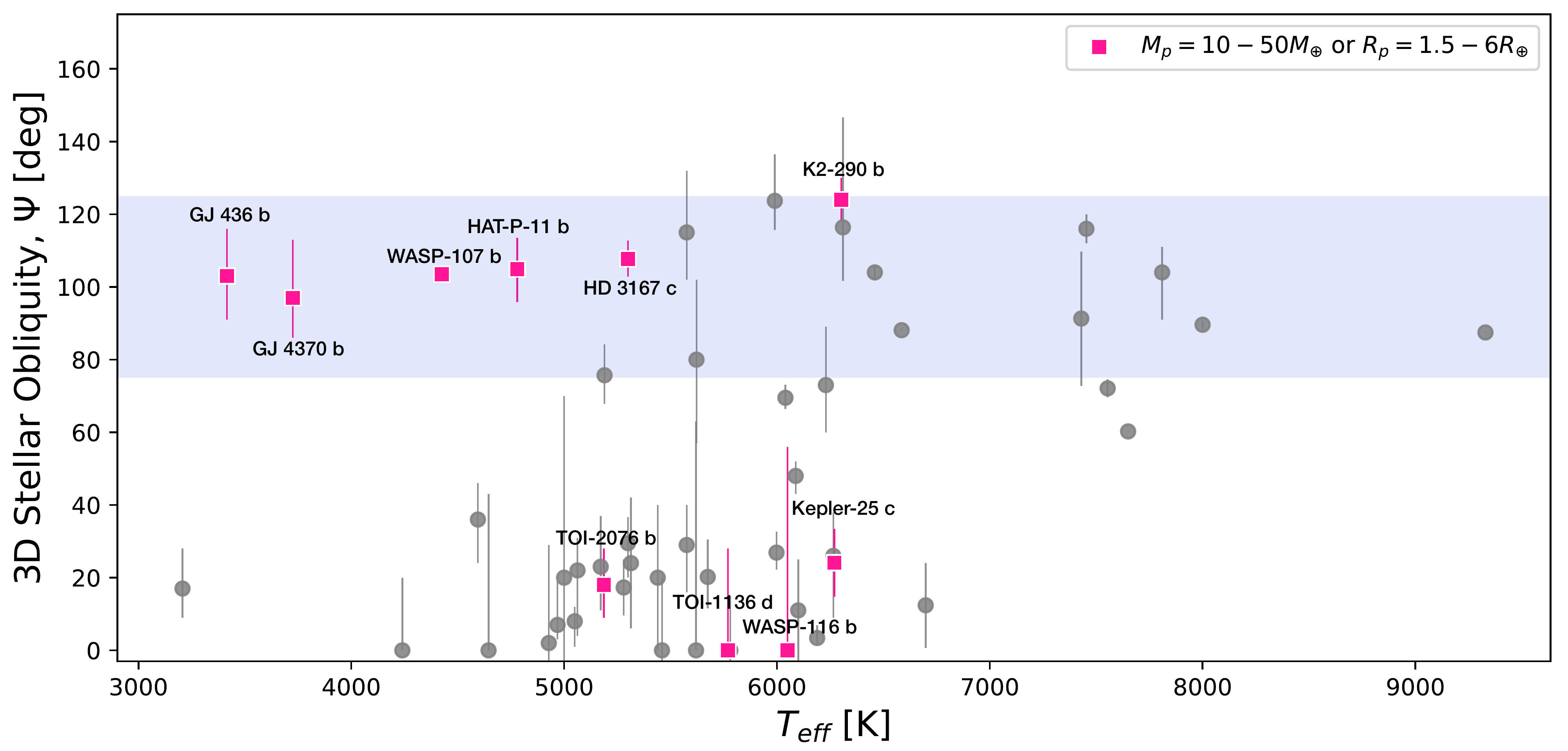}
    \caption{The 3D stellar obliquity, $\Psi$, of all planets with observed Rossiter-McLaughlin measurements, based on data from TEPCat \citep{2011MNRAS.417.2166S}. The distribution appears bimodal, consisting of planets with orbits nearly aligned or nearly perpendicular ($\Psi \in [75^\circ, 125^\circ]$; highlighted by the blue background color) with respect to their host star's equatorial planes. The pink squares represent Neptune-sized planets, which are the focus of this paper.}
    \label{fig:bimodal psi}
\end{figure*}

\citet{2021ApJ...916L...1A} compiled a sample of planets with measurements of the 3D stellar obliquities ($\Psi$), derived from sky-projected obliquities ($\lambda$) mostly measured via the Rossiter-McLaughlin effect and line-of-sight inclinations of stellar spin axes ($i_\star$). Based on their analysis, they identified two distinct planet populations: one with aligned or nearly aligned orbits, and the other with nearly perpendicular orbits with $\Psi$ falling within the range of $75^\circ - 125^\circ$ (so-called  ``polar planets''). This pattern, shown in Figure \ref{fig:bimodal psi}, has sparked discussion about whether the apparent polar planet population is statistically anomalous. \cite{2023ApJ...950L...2S} and \cite{2023AJ....166..112D} found no strong evidence for a statistical overabundance of polar planets. However, \citet{2024A&A...690A.379K} revisited this for specific categories of planets and found a preference for polar orbits at least in the subgroup of Neptunes/sub-Saturns orbiting cooler stars. Ultimately, firm conclusions on the prevalence of polar planets await confirmation from larger sample sizes. 

Nevertheless, the observed polar planets still exhibit puzzling characteristics. Specifically, the polar Neptune-sized planets (pink squares in Figure \ref{fig:bimodal psi}) are often found with moderate eccentricities ($0.03 \lesssim \rm{e} \lesssim 0.3$), puffy atmospheres, and sometimes a massive exterior companion. Examples include WASP-107~b \citep{2017A&A...604A.110A}, HAT-P-11 b \citep{2010ApJ...710.1724B}, GJ 3470 b \citep{2013ApJ...768..154D}, HD 3167 c \citep{2016ApJ...829L...9V}, K2-290 b \citep{2019MNRAS.484.3522H}, and GJ 436 b \citep{2004ApJ...617..580B}. The polar Neptunes have drawn significant attention from population studies such as the DREAM program \citep{2023A&A...669A..63B, 2023A&A...674A.120A, 2023A&A...676A.130G}.

Various theories have been proposed for generating planets on polar orbits, including a secular resonance encounter with a distant giant planet \citep{2020ApJ...902L...5P, 2024ApJ...974..304L} and von Zeipel-Lidov-Kozai oscillations \citep{2023ApJ...943L..13V}. Both of these mechanisms predict that substantial eccentricities can be generated in the process. If the polar planets attained eccentric orbits during their formation, we would expect this to generate tidal dissipation. Tidal heating is driven by time-varying distortions in the planet's shape, which are particularly pronounced in close-in planets, where tidal forces are stronger \citep{1981A&A....99..126H}. The deformations drive internal friction within the planet, converting orbital energy into heat \citep{2008ApJ...678.1396J, 2010A&A...516A..64L}. Tidal dissipation affects both the orbit and interior, and it can substantially inflate a planet's atmosphere \citep[e.g.][]{2001ApJ...548..466B, 2007A&A...462L...5L}. For planets in the Neptune to sub-Saturn regime, the radius can increase by up to a factor of $\sim2$ \citep{Millholland2019, Millholland2020}. 

In this work, we explore the hypothesis that the distinct evolutionary pathways of the aligned and misaligned (specifically polar) planets leave imprints on their physical characteristics through tidal heating. We study the observed sample of Neptune to sub-Saturn-sized planets with spin-orbit misalignment measurements and use interior structure models that account for tides to measure the degree of tidally-induced radius inflation in the aligned and misaligned groups. By examining the differences between these two groups, we aim to better understand the origins of the unusual properties of polar planets. Additionally, we conduct a case study on one of the most inflated planets in our sample, WASP-107~b. Using a simple model that couples the planet's orbital and radius evolution, we demonstrate that incorporating tidally induced radius evolution can alter a planet's dynamical history, particularly in cases of substantial radius inflation. Our findings emphasize the need for more sophisticated models that fully integrate the coupled orbital and radius evolution of a planet.

\begin{table*}[!t]
\centering
\label{tab:planet sample}
\begin{tabular}{cccccccc}
\toprule
\multicolumn{8}{c}{\textbf{Misaligned Planets}}\\
\hline
Planet    & $M_p  (M_{\oplus})$       & $R_p (R_\oplus)$     & $T_{\rm eff}$ (K)   & Eccentricity               & Semi-major axis (AU)                 & $\lambda$ ($^\circ$) & $\Psi$ ($^\circ$)  \\
\hline 
GJ 436 b    & 22.1 ± 2.3         & 4.17 ± 0.168   & 3586±36 & 0.13827 ± 0.00018 & 0.0291 ± 0.0015   & $114^{+23}_{-17}$    & $103^{+13}_{-12}$   \\
GJ 3470 b   & 13.9 ± 1.5         & 4.57 ± 0.18    & 3600±100      & 0.017 ± 0.016     & 0.0355 ± 0.0019   & $101^{+29}_{-14}$  & $97^{+16}_{-11}$    \\
HAT-P-11 b~ & 26.70 ± 2.22 & 4.36 ± 0.06    & 4780±50       & 0.218 ± 0.034     & 0.05254 ± 0.00064 & $133.9^{+7.1}_{-8.3}$  & $104.9^{+8.6}_{-9.1}$ \\
HD 3167 c   & 11.13 ± 0.78       & 3.0 ± 0.45     & 5261±60       & 0.06              & 0.1776 ± 0.0037   & $-108.9^{+5.4}_{-5.5}$ & $107.7^{+5.1}_{-4.9}$ \\
K2-290 b & 10.9 ± 3.4         & 3.06 ± 0.16    & 6302±120      & ...               & 0.0923 ± 0.0066   & $173^{+45}_{-53} $    & $124 \pm 6$   \\
WASP-107~b  & 30.5 ± 1.7         & 10.536 ± 0.224 & 4425±70       & 0.06 ± 0.04       & 0.055 ± 0.001      & $-158^{+15}_{-19}$  & $103.5^{+1.79}_{-1.8}$ \\
HAT-P-12 b  & 67.06 ± 3.81   & 10.749 ± 0.325   & 4650±60       & ...               & 0.0384 ± 0.0003  & $-54^{+41}_{-13}$  & ...   \\
HAT-P-18 b  & 62.61 ± 4.13        & 11.153 ± 0.583   & 4803±80       & 0.084 ± 0.048       & 0.0559 ± 0.0007     & 132 ± 15  & ...   \\
K2-105 b    & 30.0 ± 19.0          & 3.59 ± 0.11      & 5636±50    & ...               & 0.0833 ± 0.0039     & $-81^{+50}_{-47}$     & ...   \\
WASP-156 b  & 40.68 ± 3.18    & 5.717 ± 0.224    & 4910±61       & 0.007             & 0.0453 ± 0.0009     & $106 \pm 14$ & ...  \\   
HATS-38 b & 20.7 ± 4.8 & 6.7 ± 0.2 & 5662 ± 80 & 0.112 ± 0.072 & 0.051 ± 0.002 & $-108^{+11}_{-16}$ & ...\\
WASP-139 b	& 37.6 ±  5.5	& 8.8 ± 0.1	& 5233±60	&0.103 ±  0.050	& 0.061 ± 0.001	& $-85.6^{+7.7}_{-4.2}$	& ... \\

\hline
\multicolumn{8}{c}{\textbf{Aligned Planets}}\\
\hline
Kepler-25 c & 15.2 ± 1.3 & 5.217 ± 0.07  & 6354±27 & 0.0061 ± 0.0049 & 0.1101          & $-0.9^{+7.7}_{-6.4}$ & $24.1^{+9.29}_{-9.3}$ \\
TOI-1136 d  & 19.6 ± 7.3 & 4.627 ± 0.077 & 5770±50 & 0.016 ± 0.013   & 0.1062±0.0008    & $5 \pm 5$  &  $0^{+28}_{-0}$  \\
TOI-2076 b & 7.5 ± 2.3  & 2.518 ± 0.036 & 5200±70 & ...             & 0.0682 ± 0.0013 & $-3^{+15}_{-16}$    & $18^{+10}_{-9}$   \\
WASP-166 b  & 32.1 ± 1.6 & 7.062 ± 0.336 & 6050±50 & ...             & 0.0641 ± 0.0011 & $-0.7 \pm 0.6$  & $0^{+56}_{-0}$   \\
TOI-5398 b &  58.7 ± 5.7 & 10.30 ± 0.40 & 6000 ± 75 & $\le 0.11$ & 0.0980 ± 0.0050 & 3.0$^{+6.8}_{-4.2}$ & 13.2 ± 8.2 \\
AU Mic b  & 20.12 ± 1.57 & 4.07 ± 0.17     & 3665±31 & 0.186 ± 0.036     & 0.0645 ± 0.0013   & $-4.7^{+6.8}_{-6.4}$  & ...  \\
HD 63433 b  & $37.3 \pm 9.6$       & 2.15 ± 0.1      & 5688±28 & ...             & 0.0719 ± 0.0031   & $8^{+33}_{-45}$    & ... \\
HD 106315 c & 15.2 ± 3.7   & 4.35 ± 0.23     & 6327±48 & 0.22 ± 0.15       & 0.1536 ± 0.0017   & $-10.9^{+3.6}_{-3.8}$ & ... \\
Kepler-9 b  & 43.4 ± 1.6   & 8.29 ± 0.54     & 5774±60 & 0.0609 ± 0.001    & 0.143 ± 0.007     & $-13$ ± 16   & ... \\
Kepler-30 b & 11.3 ± 1.4   & 3.9 ± 0.2       & 5498±54 & 0.042 ± 0.003     & 0.18            & 4 ± 10    & ... \\
K2-261 b & 70 ± 10 & 9.5 ± 0.3 & 5537 ± 71 & 0.274 ± 0.065 & 0.1086±0.0012 & $32^{+48}_{-33}$ & ...\\
HD 191939 b & 10.0 $\pm$ 0.7 & 3.410±0.075 & 5348±100 & 0.031 $\pm 0.011$ & 0.0804 $\pm$ 0.0025 & $3.7 \pm 5$ & ...\\

\hline
\hline
\end{tabular}
\caption{Catalog of misaligned and aligned planets studied in this paper. The data is taken from TEPCat \citep{2011MNRAS.417.2166S} and \citet{2014AcA....64..323M, 2018Natur.553..477B, 2016MNRAS.463.2574A, 2023A&A...669A..63B, 2018AJ....155..255Y, 2024A&A...686A.127B, 2023A&A...677A..33B, 2019MNRAS.484.3522H, 2021PNAS..11817418H, 2019AJ....157..145M, 2021AJ....161...70P, 2019MNRAS.486.2290O, 2009ApJ...706..785H, 2017A&A...602A.107B, 2011ApJ...726...52H, 2014A&A...564L..13E, 2017PASJ...69...29N, 2018A&A...610A..63D, 2020AJ....160..222J, 2017MNRAS.465.3693H, 2024AJ....168..185E, 2023AJ....165...33D, 2022A&A...664A.156O, 2019MNRAS.488.3067H, 2023AJ....166..112D, 2020ApJ...899L..13H, 2021AJ....162..295C, 2024AJ....167...54C, 2024ApJS..272...32P, 2017A&A...608A..25B, 2018AJ....156...93Z, 2021ApJ...916L...1A, 2018AJ....155...70W, 2012Natur.487..449S, 2024A&A...690A.379K, 2023AJ....165..155T, 2024A&A...684L..17M, 2024AJ....168..196L}}
\end{table*}

The layout of the paper is as follows. In Section \ref{sec:planet sample}, we describe the construction of a catalog of aligned and misaligned planets used in this study. Section \ref{sec: methods} details the methodologies, focusing on two planet evolution models- with and without tides. In Section \ref{sec:pop analysis}, we perform a population analysis, applying these models to aligned and misaligned planets to find their envelope mass fractions and quantify the degree of tidally-induced radius inflation in them. Section \ref{sec:stat} assesses whether radius inflation due to tidal heating is more prevalent among misaligned planets than among aligned ones and validates the findings with statistical tests. In Section \ref{sec:wasp107}, we conduct a dynamical study of WASP 107 b, a polar planet with intriguing properties, using secular models coupled with radius evolution before finally concluding in Section \ref{sec:conclusions}.
\section{Planet Sample} 
\label{sec:planet sample}

We begin by constructing a planet sample that will enable us to perform a population-level comparative analysis between aligned and misaligned planets. We first select all planets with obliquity measurements based on data from TEPCat as of April 2025 \citep{2011MNRAS.417.2166S}. We use measurements of their physical and orbital properties available in the literature or directly from NASA Exoplanet Archive \citep{2013PASP..125..989A}. We then narrow our sample to Neptune-sized and sub-Saturn-sized planets. Specifically, we include Neptune-sized planets with masses in the range of ${M_p = 10-50 \ M_{\oplus}}$ or with radii in the range of ${R_p = 1.5-6 \ R_{\oplus}}$, and sub-Saturn-sized planets with masses in the range $M_p = 50-70 \ M_{\oplus}$. 
As a result, our analysis exclusively focuses on Neptune-sized and sub-Saturn-sized planets, deliberately excluding Jupiters. This exclusion is motivated by the distinct formation mechanisms of Jupiters, which render comparisons with Neptune- and sub-Saturn-sized planets inappropriate for the objectives of this study. To align with the scope of the analysis in this study, which utilizes simulation results from \citet{Millholland2020} (detailed in Sections \ref{sec: thermal evolution}, \ref{sec: mcmc}), we further restrict our sample to planets that have incident stellar radiation flux with respect to the Earth's, $F/F_{\oplus}$, within the range of $\log_{10}F/F_\oplus \in (0,3)$.

In addition, we exclude a few planets from our sample for the following reasons: TOI-813 b lacks reliable mass measurements; TOI-1842 b and HD 148193 b were removed due to limitations in the domain of the simulations used in this analysis (see Section \ref{sec:f_env} for details); and Pi Men c and Kepler-1656 b were excluded because the model fits (to be discussed in Section \ref{sec: thermal evolution}) failed to converge, likely due to ongoing photoevaporation causing unstable atmospheres.

The planets in the sample, selected based on the criteria outlined above, are classified as aligned or misaligned according to their measured sky-projected stellar obliquity, $\lambda$, and true obliquity, $\Psi$, where available. Specifically, planets with $|\lambda| > 40^\circ$ are classified as misaligned, while those with $|\lambda| < 40^\circ$ are classified as aligned. For the subset with measured 3D obliquities ($\Psi$), planets with $75^\circ < \Psi < 125^\circ$ are categorized as polar planets, and those with $\Psi < 25^\circ$ are classified as coplanar. In total, our sample includes 12 misaligned planets, of which 6 are identified as polar; and 12 aligned planets, with 5 classified as coplanar. The detailed classification and properties of these planets are presented in Table \ref{tab:planet sample}. For the majority of the analyses in this paper, we consider the aligned vs. misaligned classification only. 

We note that the mean radii of the misaligned and aligned planets in our sample are ${6.37 \pm 3.00 \ R_\oplus}$ and ${5.45 \pm 2.59 \ R_\oplus}$, respectively, while the mean masses are ${31 \pm 18 \ M_\oplus}$ and ${28 \pm 19 \ M_\oplus}$. There is a modest difference in average radius that is well within $1\sigma$. In the Neptune to sub-Saturn regime relevant to our study, planetary radius 
is most sensitive to the planet's envelope mass fraction \citep{2014ApJ...792....1L}, and the strength of tidal heating \citep{Millholland2019}. The slightly larger radii of misaligned planets could thus be indicative of enhanced tidal inflation. While this interpretation is preliminary, it motivates a rigorous statistical analysis, which we pursue in the following sections.


\section{Methods} \label{sec: methods}
We aim to investigate whether the misaligned planets may be experiencing different degrees of tidal dissipation and radius inflation than the aligned planets. Here we quantify the degree of tidal inflation for both aligned and misaligned populations, enabling a comparative analysis. We use the publicly available planet evolution model from \citet{Millholland2019} and \citet{Millholland2020}, which is based on a sub-Neptune evolutionary framework developed by \citet{2016Chen}.
We provide a brief overview of the tidal heating model (Section \ref{sec:tidal heating}), the thermal evolution planetary model (Section \ref{sec: thermal evolution}), and the planet parameter estimation routine (Section \ref{sec: mcmc}).

\subsection{Tidal Heating} \label{sec:tidal heating}
Tidal heating can significantly inflate a planet's radius. However, the complexity of tidal processes makes it challenging to quantify tidal dissipation and understand how the energy is distributed inside the planet. Here we adopt the viscous approach to the traditional equilibrium tide theory \citep{1966Icar....5..375G, 1981A&A....99..126H}. This approach assumes that the tidal forces exerted by the host star create an equilibrium deformation in the planet, known as the ``tidal bulge''. An imaginary line along the bulge of the planet lags the line connecting the center of the star and the planet by a constant time offset. Within this framework, the physics of tidal deformation are encapsulated in the ``reduced tidal quality factor'', $Q' \equiv 3Q/2k_2$, where $Q$ is the tidal quality factor and $k_2$ is the planet's Love number. The exact value of  $Q'$ is highly uncertain for any given planet, so we explore a wide range of plausible values in our analysis. 

The tidal luminosity (i.e. the rate at which the planet's orbital energy is converted into heat energy through tidal dissipation) is given by \cite{2010A&A...516A..64L} as follows:
\begin{equation} \label{eq:L_tide}
    L_{\rm tide}(e,\epsilon) = 2K\left[N_{\rm a}(e) - \frac{N^2(e)}{\Omega(e)}\frac{2\rm cos^2(\epsilon)}{1 + \rm cos^2(\epsilon)} \right],
\end{equation}
where $e$ is the eccentricity, $\epsilon$ is the planetary obliquity, and $N_{\rm a}(e), N(e)$, and $\Omega(e)$ are defined as
\begin{equation} \label{n_a}
    N_{\rm a}(e) = \frac{1 + \frac{31}{2}e^2 + \frac{255}{8}e^4 + \frac{185}{16}e^6 + \frac{25}{64}e^8}{(1-e^2)^{\frac{15}{2}}}
\end{equation}
\begin{equation} \label{n(e)}
    N(e) = \frac{1 + \frac{15}{2}e^2 + \frac{45}{8}e^4 + \frac{5}{16}e^6}{(1-e^2)^6}
\end{equation}
\begin{equation}
    \Omega(e) = \frac{1 + 3e^2 + \frac{3}{8}e^4}{(1 - e^2)^{\frac{9}{2}}}.
\end{equation}
The quantity $K$ is the characteristic luminosity scale given by
\begin{equation} \label{K}
    K = \frac{9n}{4Q'}\left(\frac{GM_{\star}^2}{R_p}\right) \left(\frac{R_p}{a}\right)^6
\end{equation}
where $M_{\star}$ is the stellar mass, $R_p$ is the planet radius, $a$ is the semi-major axis, and  $n = 2\pi/P$ is the mean motion. Finally, equation \ref{eq:L_tide} also assumes that the planet's spin rotation frequency $\omega = 2\pi/P_{\rm rot}$ has reached an equilibrium ($d\omega/dt = 0$) and is given by \citep{2010A&A...516A..64L}:
\begin{equation}
    \omega_{\rm eq} = n\frac{N(e)}{\Omega(e)}\frac{2\rm cos\epsilon}{1+\rm cos^2\epsilon}.
    \label{eq: omega_eq}
\end{equation} 

Before proceeding, we note that tides are not the only potential source of internal heating that could cause radius inflation. Another possibility is ohmic dissipation, which arises in planets with weakly ionized atmospheric winds and magnetic fields, where the motion of charged particles across magnetic field lines induces electric currents. These currents can deposit energy deep in the interior by removing kinetic energy from the atmospheric flow \citep{2008Icar..196..653L, 2011ApJ...738....1B}. To assess its potential contribution, we estimate the ohmic luminosity ($L_{\rm Ohm}$), the power deposited via ohmic heating, using the fitting relations from \citet{2017ApJ...846...47P}. Although these relations are approximate, they provide useful order-of-magnitude estimates. We find that $L_{\rm Ohm}$ is on average $\sim10^4$ times smaller than $L_{\rm tide}$ across our planet sample (where the details of the $L_{\rm tide}$ calculation are described later in Section \ref{sec:pop analysis} and Table \ref{tab:radius inflation}). This disparity suggests that ohmic heating is generally subdominant compared to tidal dissipation, and we therefore exclude it from our models. However, we note that in a small number of systems, 
$L_{\rm Ohm}$ can be comparable to $L_{\rm tide}$ \citep[e.g.][]{2025ApJ...985...87B}, in which cases ohmic heating could be incorporated into more detailed individual planet models in the future. Given that our study focuses on population-level trends, we consider its omission justified in this context.
 
\subsection{Thermal Evolution Model} \label{sec: thermal evolution}
 The planetary thermal evolution model from \citet{Millholland2019, Millholland2020} assumes a spherically symmetric, two-layer planetary structure comprised of a dense core of heavy elements and a surrounding H/He-dominated envelope. The thermal and structural evolution of the atmosphere is simulated using the Modules for Experiments in Stellar Astrophysics (MESA; \citealt{2011ApJS..192....3P, 2013ApJS..208....4P, 2015ApJS..220...15P, 2018ApJS..234...34P}) 1D stellar evolution code, adapted with modifications introduced by \citet{2016Chen}. The tidal luminosity is incorporated as an additional component in the core luminosity. Additional details of the model are provided in \citet{Millholland2019, Millholland2020}.
 
 We adopt the simulation suite from \citet{Millholland2019, Millholland2020} for this study. This consists of a set of $\sim15,000$ planetary models, varied randomly in four principal parameters given in Table \ref{tab:parameters}: the planet mass $M_p$, the fraction of mass in the H/He envelope $f_{\rm env}$, the strength of incident stellar radiation flux with respect to the Earth's $F/F_{\oplus}$, and the strength of tidal dissipation $\Gamma$ which is defined as
\begin{equation} \label{gamma}
    \Gamma = \log_{10}\left[{\frac{Q'(1+\cos^2{\epsilon})}{\sin^2{\epsilon}}}\right]
\end{equation} 
where $\epsilon$ is the planetary obliquity. 
For each set of parameters, there are two simulations, one including tidal heating and the other neglecting tides with no dependence on the tidal strength parameter. The planets are evolved for 10 Gyr in both cases.  

 \begin{table}
    \centering
    \begin{tabular}{cc}
    \toprule
        Parameter  & Range \\
        \hline
        $M_p/M_{\oplus}$ &  (1,70)\\
        $\rm log_{10}f_{\rm env}$  & (-2.5,-0.3)\\
        $\rm log_{10}F/F_{\oplus}$ &  (0,3) \\
        $\Gamma$   & (3,7) \\
\hline
\hline
    \end{tabular}
    \caption{Parameters and their ranges for the MESA simulations by \citet{Millholland2020}.}
     \label{tab:parameters}
\end{table}

\subsection{MCMC Parameter Inference} \label{sec: mcmc}
Our goal is to infer the interior structures of observed planets with an aim of determining how strongly they have been affected by tides. We do this by employing an interpolation to the simulation suite just described, coupled with a Markov Chain Monte Carlo (MCMC) inference method detailed in \citet{Millholland2020}. We study the planet's structure using two MCMC fits associated with the two models: one including tidal heating and the other neglecting tides. The MCMC approach yields the posterior distribution of planet parameters by demanding that the model radius agrees with the planet's observed radius within uncertainties. 

For the fit neglecting tides, we fix $F/F_{\oplus}$ to the observed value and allow $M_p$ to float within $3\sigma$ uncertainties about the observed value, resulting in only one effective free parameter that we wish to constrain: $f_{\rm env}$. On the other hand, the fit including tidal heating has two effective free parameters: $f_{\rm env}$ and $\Gamma$. We denote the envelope mass fractions obtained from these two models without tides and with tides as $f_{\rm env, 0}$ and $f_{\rm env, t}$, respectively. The difference between the resulting posterior distributions indicates how strongly tides have affected the fit. 

In this analysis, the planetary obliquity, for all planets, is set to $\epsilon = 0^\circ$, to isolate the effects of eccentricity tides. For planets with measured eccentricities, we use the observed values; for those without eccentricity measurements, we assume a small eccentricity of $e = 0.05 \pm 0.02$. This is a relatively conservative assumption since many planets could have larger eccentricities. However, later on in the paper, we consider the sample isolated to planets with measured eccentricities only.

The MESA simulations run by \citet{Millholland2019} and \cite{Millholland2020}, were based on $e=0$ and $\epsilon \neq 0$ (circular orbits, non-zero planetary obliquities) and solar parameters for the star. To generalize the results, we use the following transformation equation that incorporates parameters such as eccentricity, semi-major axis, planetary obliquity (set to 0 for this analysis), and stellar mass to transform $\Gamma$ into $\log_{10}Q'$. 
\begin{align}
    \label{eq:transformation}
    \log_{10}Q' = & \Gamma_i + \log_{10}\Biggr[ \left(\frac{M_\star}{M_{\star, i}}\right)\left(\frac{n}{n_i}\right)\left(\frac{a_i}{a}\right)^6 \nonumber \\
         &\times \left(N_a(e) - \frac{N^2(e)}{\Omega(e)}\right) \Biggr].
\end{align}
Here the subscript ``\textit{i}'' corresponds to the initial MESA simulation results using $e = 0$ and solar parameters. Therefore, $M_{\star, i} = M_\odot$, and $a_i/\rm {au} = (F/F_\oplus)^{(-1/2)}$. The transformation ensures that the tidal luminosity ($L_{\rm tide}$) remains invariant, such that ${L_{\rm tide}(e, \epsilon=0)=L_{\rm tide}(e_i=0,\epsilon_i)}$. By preserving this invariance, the transformation modifies the initial tidal strength, $\Gamma_i$, to a transformed $\log_{10}{Q'}$ value. This enables the posterior distributions to account for arbitrary $e$ and stellar parameters, making the results more generalizable across different planetary and stellar systems. 

\section{Population Analysis} \label{sec:pop analysis}
\begin{figure}
    \centering
    \includegraphics[width=0.42\textwidth]{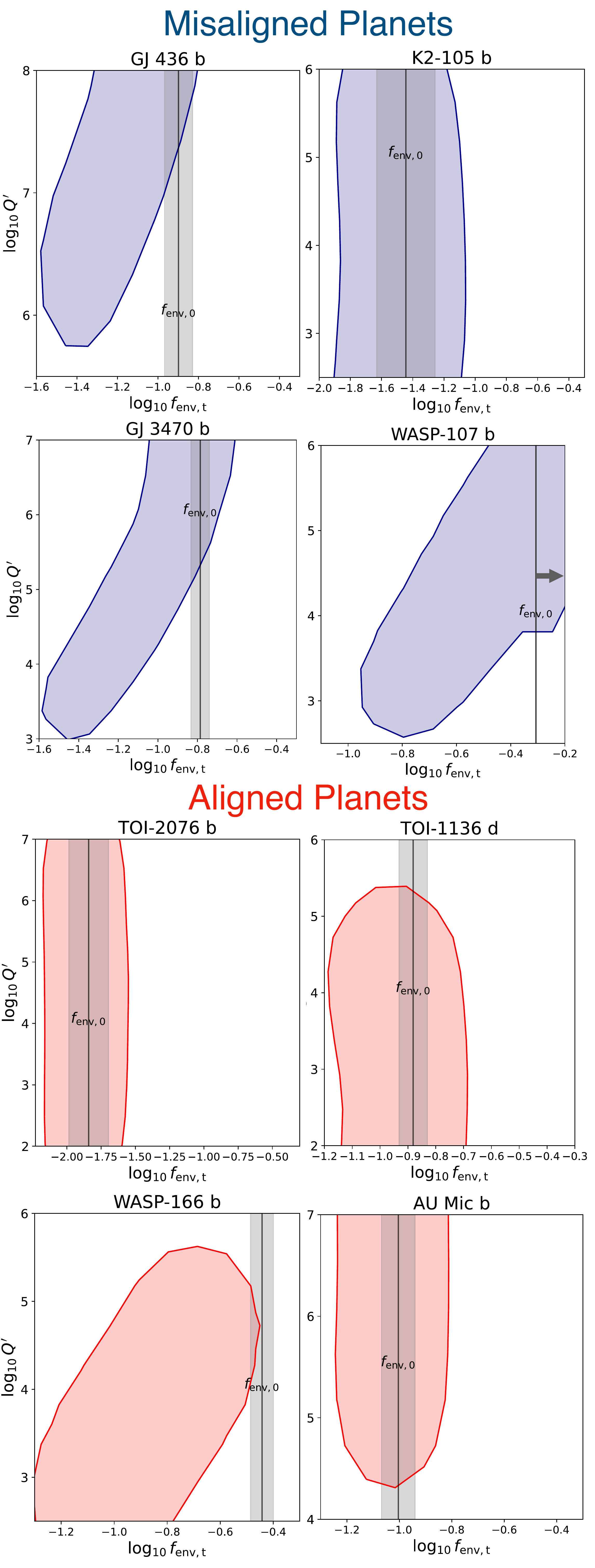}
    \caption{Examples of $f_{\rm env}$ distributions from the MCMC analysis, shown for a handful of misaligned (top panel) and aligned planets (bottom panel). In each plot, the solid gray line represents the envelope mass fraction obtained for a planet using a tides-free model ($f_{\rm env, 0}$), with the gray-shaded area indicating the $1\sigma$ uncertainty. The other shaded regions -- blue for misaligned planets and red for aligned planets -- illustrate the posterior distribution of $f_{\rm env,t}$ vs. $Q'$, derived from the model including tides.}
    \label{fig:f_env sim}
\end{figure}
We now apply the parameter inference method to the aligned and misaligned planets listed in Table \ref{tab:planet sample} to evaluate the strength of tidal dissipation and the extent of radius inflation. This comparison will allow us to evaluate how a planet's orbital history and the accompanying tidal evolution affect its physical characteristics, particularly whether the impact of tidal heating differs between the aligned and misaligned planet populations. 

\subsection{$f_{\rm env, 0}$, and $f_{\rm env, t}$ Estimation} \label{sec:f_env}
We run the two variations of the MCMC -- one excluding tides and the other including them, as outlined in Section \ref{sec: mcmc} -- for all the planets in our sample. These simulations yield the posterior distributions of $f_{\rm env, 0}$ and $f_{\rm env, t}$. Examples of $f_{\rm env}$ distributions resulting from the MCMC analysis for some aligned and misaligned planets in our sample are presented in Figure \ref{fig:f_env sim}. The plots display the estimates of the planet's envelope mass fraction using the following models:
\begin{enumerate}[label = (\roman*)]
    \item Neglecting tidal inflation ($f_{\rm env, 0}$): The $f_{\rm env, 0}$ and its error are estimated as the mean and the standard deviation (1$\sigma$) of the posterior distribution obtained from the tides-free model. These values are indicated by the gray line and the bar in the plots from Figure \ref{fig:f_env sim}.
    \item Including tidal inflation ($f_{\rm env, t}$): The blue and red regions in the plots represent the 2D posterior distributions in $\rm log_{10}Q'$ and $f_{\rm env, t}$ from the fits obtained from the model including tidal heating.
\end{enumerate}
These examples indicate that some planets are not strongly affected by tides while others are. The cases where the constraints on $f_{\rm env, t}$ and $f_{\rm env, 0}$ overlay each other are those where tides do not have a substantial impact on the fit. In contrast, the cases where the constraints on $f_{\rm env, t}$ fall well below those on $f_{\rm env, 0}$ indicate that tides have a substantial impact on the fit; the tidally-induced inflation results in a smaller inferred envelope fraction in a manner that gets more pronounced as $Q'$ decreases.

We quantify the impact of tidal heating on the envelope mass fraction estimates of all planets in our sample by comparing the values of $f_{\rm env, 0}$ and $f_{\rm env, t}$ at a fixed $Q'$ value. Maintaining a consistent $Q'$ for all planets is essential to ensure a uniform comparison. For this purpose, we fix $Q' = 10^{4}$, which is a reasonable approximation for planets in the size range analyzed here \citep{2017AJ....153...86M, 2018AJ....155..157P}. However, we note that choosing a different value would not significantly impact the comparisons between $f_{\rm env, 0}$ and $f_{\rm env, t}$.  

In some cases, the posterior distributions from the fit including tides do not extend to $Q' = 10^4$. Examples include planets such as GJ 436 b and AU Mic b, as illustrated in Figure \ref{fig:f_env sim}. This limitation arises from the transformation of $\Gamma$ into $\log_{10}Q'$ as described in Section \ref{sec: mcmc}. Since the transformation depends on $e$ and stellar parameters, which vary across our sample, the posterior distributions from MCMC fits for some planets may not extend to $Q' \approx 10^4$. To address this issue, we extrapolate the distributions of $\log_{10}{Q'}$ to the required value. We assume a linear dependence of $\log_{10}f_{\rm env, t}$ vs. $\rm \log_{10}Q'$, an assumption that appears valid based on the 2D posterior distributions obtained for our sample, as shown in Figure \ref{fig:f_env sim}. Accordingly, we fit a straight line to $\log_{10}f_{\rm env, t}$ vs. $\rm log_{10}Q'$ distribution in the region where $f_{\rm env,t}$ is less than $f_{\rm env, 0}$, since $f_{\rm env,t}$  cannot theoretically exceed $f_{\rm env, 0}$. Beyond this point, $f_{\rm env, t}$ remains constant at $f_{\rm env, 0}$. For each planet in our sample, we then calculate the average $\log_{10}f_{\rm env, t}$ associated with $\rm log_{10}Q' \in [3.8, 4.2]$ either using the fitted line or extrapolating the fitted line, as appropriate. The error is estimated as the standard deviation of $\log_{10}f_{\rm env, t}$ within this same bin. Our estimated values of $f_{\rm env, 0}$ and of $f_{\rm env, t}$ at a fixed $Q' = 10^4$ are provided in Table \ref{tab:radius inflation}. We save a discussion of these results until Section \ref{sec:stat}.

Additionally, as our models rely on the simulation results from \citet{Millholland2020}, which define the 2-D posterior distributions only up to an envelope mass fraction of 50\%, we were unable to obtain valid fits for some planets, such as TOI-1842 b and HD 148193 b. For these cases, their radii are so large that fitting with the model neglecting tides would require $f_{\rm env, 0} > 50\%$, which lies outside the bounds of the simulation grid. As a result, we were unable to quantify the impact of tidal heating for these planets, and therefore excluded them from our analysis

\begin{table*}
\centering 
\begin{tabular}{llllll}
\toprule
\multicolumn{6}{c}{\textbf{Misaligned Planets}}\\
\hline
Planet   & $f_{\rm env, 0}$ [\%]     & $f_{\rm env, t}$ [\%]  & $R_{\rm p,t}/R_{\rm p,0}$ & $L_{\rm tide}$ [erg $\rm s^{-1}$]& $\rm log_{10}(L_{\rm tide}/L_{\rm irr})$  \\
\hline
GJ 436 b    & 12.8 ± 2   & 1.02 ± 1.34  & 1.56 ± 0.21  & $1.099 \times 10^{27}$ & 0.07 ± 0.09            \\
GJ 3470 b   & 16.5 ± 1.8 & 4.77 ± 2.24  & 1.42 ± 0.27  & $7.178 \times 10^{24}$ & -2.46 ± 0.84          \\
HAT-P-11 b & 13.3 ± 1.2 & 2.25 ± 1.58 & 1.43 ± 0.22 & $2.008 \times 10^{26}$  & -1.13 ± 0.21          \\
HD 3167 c   & 4.3 ± 3.2  & 2.95 ± 1.67  & 1.02 ± 0.07 & $ 1.930 \times 10^{20}$  & -6.24 ± 0.45          \\
K2-290 b   & 3.1 ± 1  & 1.68 ± 0.70  &  1.11 ± 0.04 & $4.841 \times 10^{22}$   & -5.21 ± 0.57           \\
WASP-107~b  & $>50$ & 24.19 ± 4.33 & 1.89 ± 0.47 & $3.996 \times 10^{26}$  & -1.59 ± 0.79          \\
HAT-P-12 b  & $>50$ & 24.47 ± 3.66  & 1.73 ± 0.25 & $5.387 \times 10^{27}$   & -0.81 ± 0.50             \\
HAT-P-18 b  & $>50$   & 35.68 ± 5.16  & 1.68 ± 0.27 & $1.292 \times 10^{27}$    & -1.33 ± 0.76            \\
K2-105 b    & 3.9 ± 1.7    & 3.12 ± 1.74 & 1.03 ± 0.06 & $1.728 \times 10^{23}$  & -4.26 ± 0.50             \\
WASP-156 b  & 25.1 ± 3.2   & 13.18 ± 2.05   & 1.21 ± 0.34 & $1.814 \times 10^{24}$   & -3.25 ± 0.97            \\
HATS-38 b   & 40.3 ± 2  & 11.47 ± 2.38   &  1.55 ± 0.20 & $1.243 \times 10^{26}$ & -1.92 ± 0.84             \\
WASP-139 b  & 46.2 ± 2.8  & 45.84 ± 2.90  & 1.00 ± 0.01 & $1.058 \times 10^{26}$                & -1.72 ± 0.66     \\
\hline
\multicolumn{6}{c}{\textbf{Aligned Planets}}\\
\hline
Kepler-25 c & 22 ± 1  & 19.39 ± 2.08 & 1.10 ± 0.10 & $2.474 \times 10^{21}$  & -6.82 ± 0.87  \\
TOI-1136 d  & 13.2 ± 1.5 & 11.22 ± 2.07 &  1.05 ± 0.08  & $3.305 \times 10^{21}$ & -5.72 ± 0.83  \\
TOI-2076 b & 1.5 ± 0.5  & 1.36 ± 0.46   & 1.01 ± 0.02 & $5.094 \times 10^{22}$        & -4.81 ± 0.54     \\
WASP-166 b  & 36.3 ± 3.5  & 15.00 ± 3.59 & 1.39 ± 0.29 & $4.781 \times 10^{25}$   & -3.07 ± 0.50  \\
TOI-5398 b & 48.3 ± 1 & 48.06 ± 3.31 & 1.00 ± 0.01 & 6.148 $\times 10^{25}$ & -2.61 ± 0.79   \\
AU Mic b  & 10 ± 1.5   & 9.34 ± 1.47  & 1.02 ± 0.03 & $5.966 \times 10^{24}$                  & -1.98 ± 0.23           \\
HD 63433 b  & 0.6 ± 0.2   & 0.50 ± 0.13 & 1.00 ± 0.01 & $3.328 \times 10^{22}$    & -4.66 ± 0.50             \\
HD 106315 c & 14.2 ± 2.4 & 9.32 ± 2.38 & 1.16 ± 0.31 & $1.371 \times 10^{23}$    & -4.50 ± 1.05              \\
Kepler-9 b  & $>50$ & 40.04 ± 3.27 & 1.18 ± 0.24 & $2.691 \times 10^{23}$                & -4.29 ± 0.09              \\
Kepler-30 b & 11.2 ± 1.8 & 10.08 ± 1.79  & 1.03 ± 0.06 & $4.512 \times 10^{20}$    & -6.18 ± 0.15             \\
K2-261 b & 45.6 ± 9.8  & 30.57 ± 4.63 & 1.53 ± 0.32 & 1.994 $\times 10^{26}$ & -2.20 ± 0.37\\
HD 191939 b & $8.2 \pm 0.7$ & 4.79 $\pm 2.52$ & 1.17 $\pm$ 0.11 & 4.17 $\times 10^{22}$ & -4.72 $\pm$ 0.36 \\
\hline
\hline
\end{tabular}
\caption{Estimates of the envelope mass fraction obtained using models with tides ($f_{\rm env, t}$) and without tides ($f_{\rm env, 0}$), the degree of inflation ($R_{\rm p,t}/R_{\rm p,0}$), and tidal luminosity ($L_{\rm tide}$), all calculated at $Q' = 10^4$. These values were derived following the procedures described in Sections \ref{sec:f_env} \& \ref{sec: radius estimation} and include all aligned and misaligned planets cataloged in Table \ref{tab:planet sample}. We assumed $\epsilon = 0^\circ$ and used measured eccentricities for planets with observed values; for those without eccentricity measurements, or with $e = 0$, we assumed a small nominal eccentricity of $e = 0.05 \pm 0.02$.} \label{tab:radius inflation}
\end{table*}

\subsection{$R_{\rm p, 0}$, and $R_{\rm p, t}$ Estimation} \label{sec: radius estimation}
\begin{figure}
    \centering
    \includegraphics[width=0.5\textwidth]{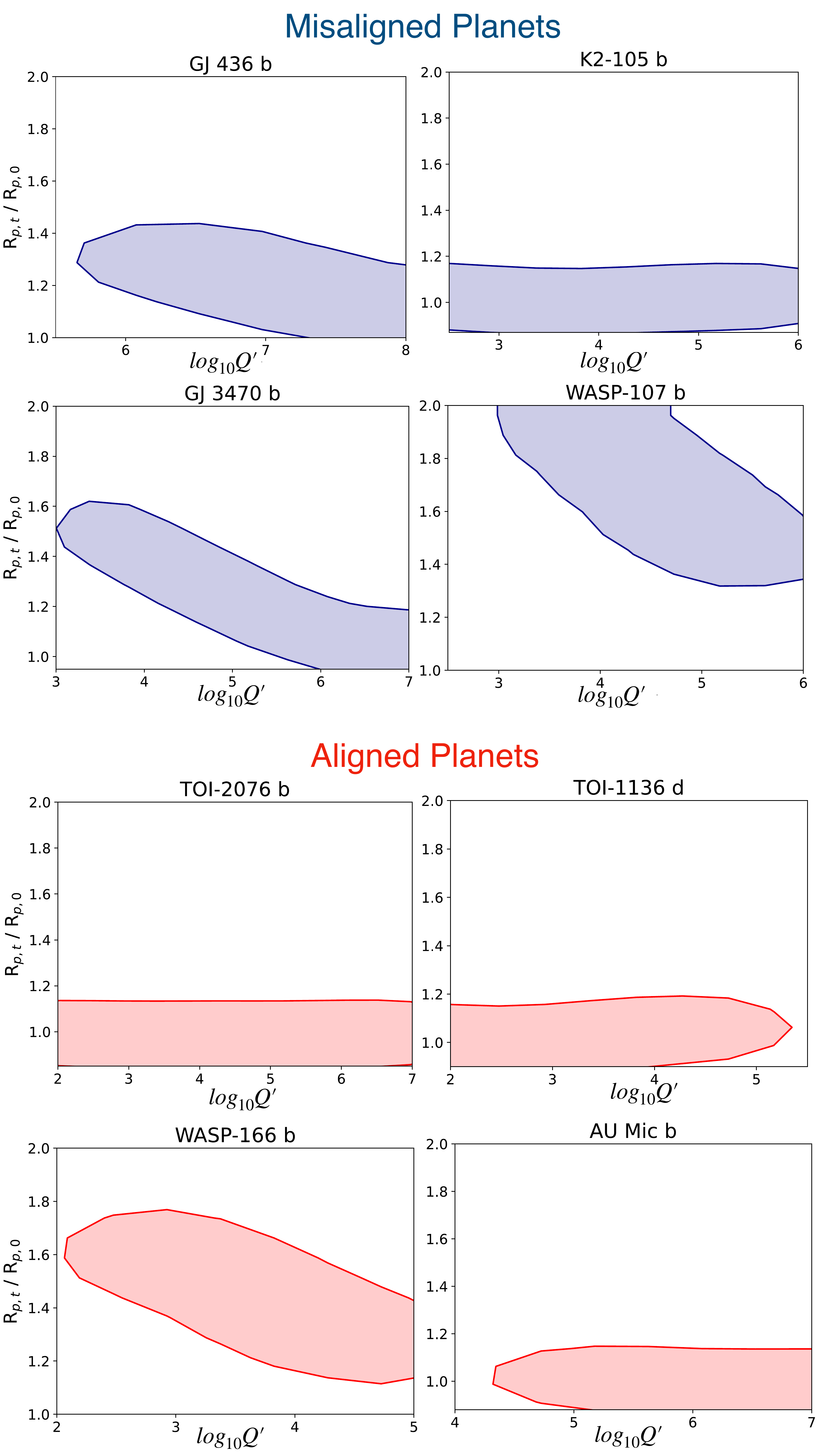}
    \caption{Examples of the posterior distributions of the degree of inflation, $R_{p,t}/R_{p,0}$, as a function of $Q'$. The top and bottom panels show examples for misaligned planets and aligned planets, respectively.}
    \label{fig:R_pt/RP0}
\end{figure}
In addition to calculating $f_{\rm env, t}$ and $f_{\rm env, 0}$, it is useful to compute a measure of the impact of tidal heating on the planet's radius. We use $f_{\rm env, t}$ from the model including tides to compute a radius we define as $R_{p,0}$, which represents the radius the planet would have if its envelope mass fraction was equal to $f_{\rm env, t}$ but it was not inflated by tidal heating. This can be compared to $R_{p,t}$, which represents the radius associated with $f_{\rm env, t}$ and the tidally heated model. (Note $R_{p,t}$ equals the planet's observed radius by construction of the model.) We define the ratio $R_{p,t}/R_{p,0} \geq 1$ as the ``degree of inflation''. The larger this is, the more pronounced the impacts of tidal heating on inflation, with $R_{p,t}/R_{p,0} = 1$ corresponding to no effect. The constraints on $R_{p,t}/ R_{p,0}$ vs. $\log_{10}Q'$ for a handful of planets in our sample are shown in Figure \ref{fig:R_pt/RP0}. As expected, the results agree with the envelope fraction constraints in Figure \ref{fig:f_env sim}.

To quantify the degree of inflation at a fixed ${\log_{10}Q'=4}$, we adopt the methodology outlined in Section \ref{sec:f_env} for determining $f_{\rm env, t}$ at a fixed $Q'$. We assume a linear dependence of $R_{\rm p,t}/R_{\rm p,0}$ vs. $\rm log_{10}Q'$ but only for the range where $R_{\rm p,t}/R_{\rm p,0} > 1$. This linear trend is evident in the examples shown in Figure \ref{fig:R_pt/RP0}. Once $R_{\rm p,t}/R_{\rm p,0}$ reaches a value of 1, it remains constant, as it cannot drop below 1. For $\log_{10}Q'$ values where $R_{\rm p,t}/R_{\rm p,0} > 1$, we fit a straight line to the $R_{\rm p,t}/R_{\rm p,0}$ vs. $\rm log_{10}Q'$ plot. We calculate the average $R_{\rm p,t}/R_{\rm p,0}$ associated with $\rm log_{10}Q' \in [3.8, 4.2]$ using the fitted line for each planet. If the posterior distributions do not extend to this range of $\log_{10}Q'$ values, we linearly extrapolate the fitted line to estimate the values. The associated error is estimated as the standard deviation of the $R_{\rm p,t}/R_{\rm p,0}$ within the same bin. Our estimated values of the degree of inflation, $R_{\rm p,t}/R_{\rm p,0}$, at fixed $Q' = 10^4$ are provided in Table \ref{tab:radius inflation}. We save a discussion of these results until the next section.

\section{Are Misaligned Planets Predominantly Puffier?} \label{sec:stat}
\subsection{Reduction of $f_{\rm env}$ When Accounting for Tides}

\begin{figure*}
    \centering
    \includegraphics[width=0.95\textwidth]{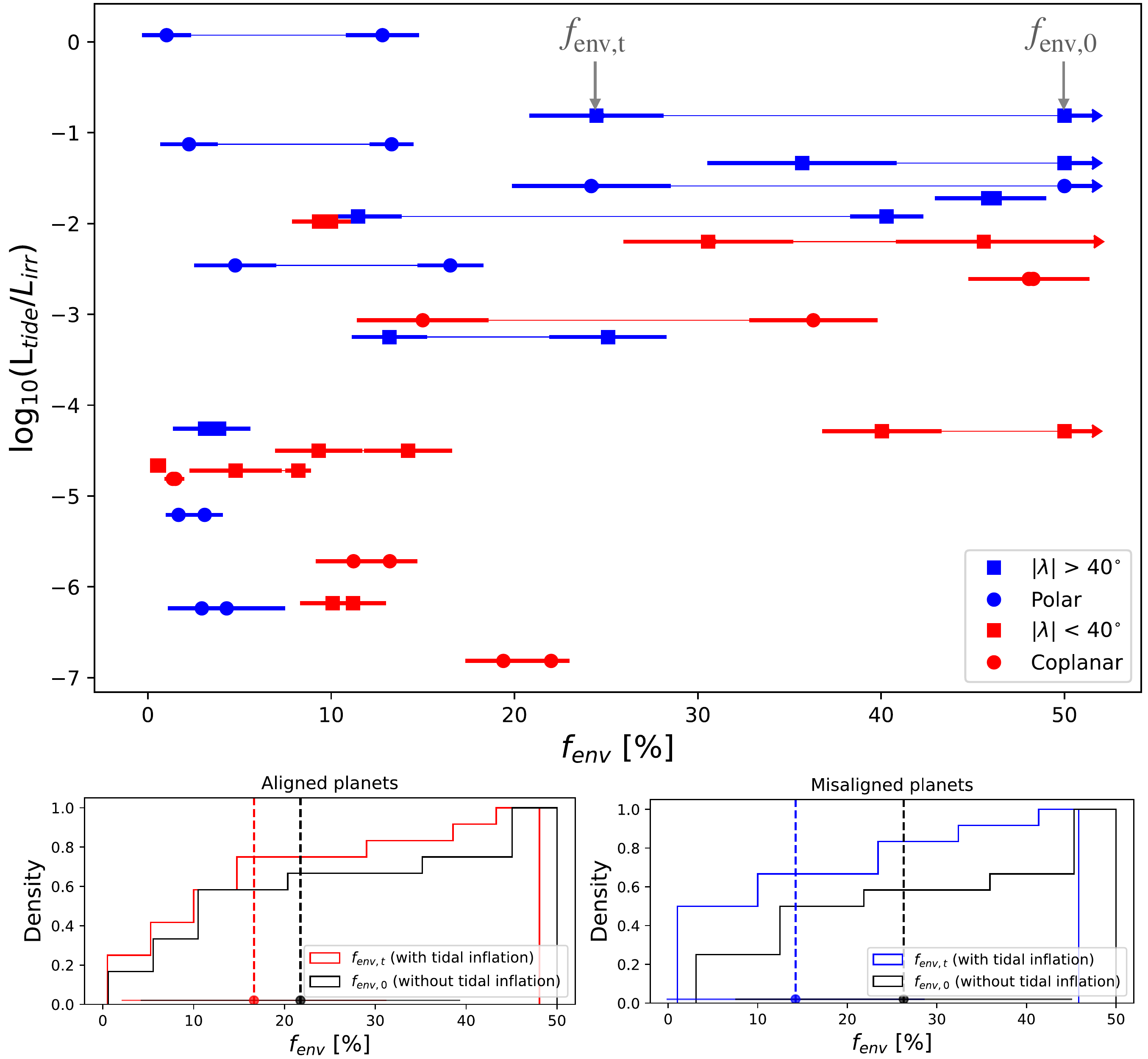}
    \caption{Impact of of tidal inflation on the estimated envelope mass fraction. The top panel illustrates the reduction in the estimated envelope mass fraction, from $f_{\rm env, 0}$ (without tides) to $f_{\rm env, t}$ (including tides at $Q' = 10^4$). Aligned and misaligned planets are depicted with red and blue squares and dots, respectively. The bottom left and right plots represent the CDFs of $f_{\rm env, t}$ and $f_{\rm env, 0}$ across the sample of aligned planets and misaligned planets, respectively. The reduction in $f_{\rm env}$ when tides are included in the model is clearly more significant for the misaligned planet population.}
    \label{fig:f_env}
\end{figure*}

We first compare the envelope fraction estimates from the models with and without tides. As shown in the top panel of Figure \ref{fig:f_env}, $f_{\rm env, 0} \ge f_{\rm env, t}$, resulting in $f_{\rm env, 0} - f_{\rm env, t} \ge 0$ for all planets in our sample. Moreover, this difference increases with higher values of $\rm log_{10}(L_{\rm tide}/L_{\rm irr})$, where $L_{\rm tide}$ is the tidal luminosity (as defined in equation \ref{eq:L_tide} and assuming $Q' = 10^4$) and $L_{\rm irr}$ is the incident stellar power. This behavior arises because tidal heating causes planets to expand to larger radii for a given $f_{\rm env}$. As a result, the tidal model predicts lower $f_{\rm env}$ values to explain a planet's observed radius compared to the tides-free model, as was also seen in Figure \ref{fig:f_env sim}. A larger $\rm log_{10}(L_{\rm tide}/L_{\rm irr})$ indicates stronger tidal heating, which explains the increasing difference between $f_{\rm env, 0}$ and $f_{\rm env, t}$ for larger $\rm log_{10}(L_{\rm tide}/L_{\rm irr})$ values. Across the entire sample, the mean difference is $\langle f_{\rm env, 0} - f_{\rm env, t}\rangle \approx 9\%$.

Figure \ref{fig:f_env} (bottom panel) shows the cumulative distribution functions (CDFs) of the envelope mass fractions for both aligned (left) and misaligned (right) planets. In the bottom left panel, it is evident that the CDFs of $f_{\rm env, t}$ and $f_{\rm env, 0}$ for aligned planets are fairly similar. In fact, the mean difference for the aligned planet population is $\langle f_{\rm env, 0} - f_{\rm env, t}\rangle _a\approx5\%$, indicating a smaller impact of tidal heating. In contrast, the bottom right panel for misaligned planets reveals a different scenario. For this population, the mean difference is $\langle f_{\rm env, 0} - f_{\rm env, t}\rangle _m\approx12\%$, more than twice the value attained for the aligned planet population. This indicates a significant reduction in the envelope fraction estimate predicted by the model that includes tides, suggesting that misaligned planets experience more tidal heating. We will evaluate this claim from multiple perspectives throughout this subsection.  

\subsection{Radius Inflation}
\begin{figure*}
    \centering
    \includegraphics[width=\textwidth]{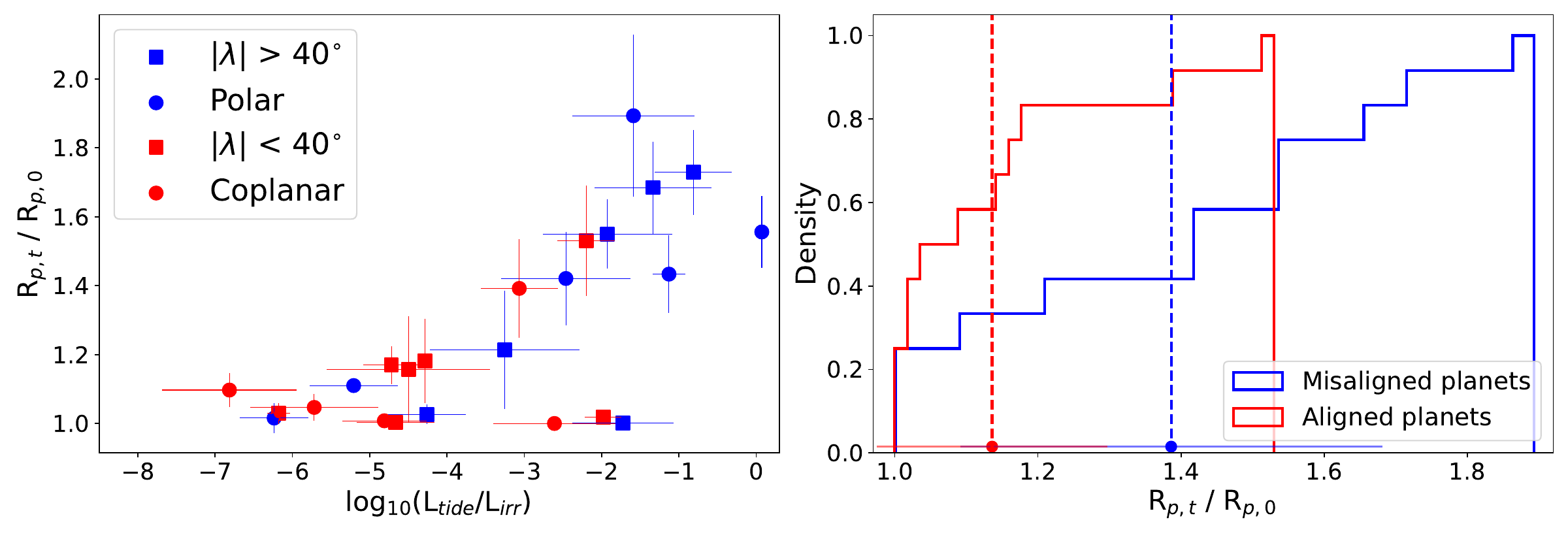}
    \caption{Summary of results on the degree of inflation, $R_{\rm p,t}/R_{\rm p,0}$. The left panel shows $R_{\rm p,t}/R_{\rm p,0}$ as a function of $\log_{10}(L_{\rm tide}/L_{\rm irr})$, calculated for our entire sample at $Q' = 10^4$. Aligned planets are represented by red dots and squares, while misaligned planets are represented by blue dots and squares. The upper right region of the plot reveals an overpopulation of misaligned planets, characterized by greater radius inflation and higher $\log_{10}(L_{\rm tide}/L_{\rm irr})$. The right panel shows the CDFs of $R_{\rm p,t}/R_{\rm p,0}$ for the aligned and misaligned planet populations, depicted by the red and blue curves, respectively. The red and blue dashed lines are the average $R_{\rm p,t}/R_{\rm p,0}$ values for aligned and misaligned planets respectively.}
    \label{fig:radius inflation}
\end{figure*}

We can also quantify the effects of tides by comparing the degree of inflation, introduced in Section \ref{sec: radius estimation}. For our sample, the average degree of inflation is $\langle R_{p,t}/R_{p,0}\rangle\approx1.26$, with values ranging from 1 to 1.9. The left panel of Figure \ref{fig:radius inflation} displays the degree of inflation as a function of $L_{\rm tide}/ L_{\rm irr}$. As anticipated, tides significantly inflate the planetary radius when $\log_{10}(L_{\rm tide}/L_{\rm irr}) \gtrsim -5$ \citep{Millholland2019}, consistent with the expectation that higher tidal luminosities correlate with greater tidal inflation. 

Additionally, Figure \ref{fig:radius inflation} shows evidence suggesting that misaligned planets (denoted by blue dots and squares) may exhibit more inflation compared to aligned ones (denoted by red dots and squares). This trend is particularly evident in the upper right region of the plot, where higher $L_{\rm tide}/L_{\rm irr}$ values and greater radius inflation are predominantly concentrated with misaligned planets. Furthermore, the average degree of inflation for the aligned planet population is $\langle R_{\rm p,t}/R_{\rm p,0}\rangle_a \approx 1.1$, indicating little to no tidal inflation. In contrast, the average for the misaligned population is $\langle R_{\rm p,t}/R_{\rm p,0}\rangle_m \approx 1.4$, suggesting more significant inflation. These averages are also represented in the right panel of Figure \ref{fig:radius inflation} by red and blue dashed lines for the aligned and misaligned planet populations, respectively. While these results are suggestive that the misaligned planets are more tidally inflated, further statistical testing is required to confirm this result.

\subsection{Statistical Testing}
To substantiate our claims and statistically identify differences in tidal inflation between aligned and misaligned planets, we perform the following statistical tests on the degree of inflation, $R_{\rm p,t}/R_{\rm p, 0}$.


\subsubsection{Kolmogorov-Smirnov Two-Sample Test}
The \textit{Kolmogorov-Smirnov (KS) test} is a nonparametric test used to determine whether two samples are consistent with being drawn from the same underlying distribution. The assessment of the \textit{K-S Test} is conducted through hypothesis testing, where the null hypothesis, $H_0$, posits that both samples are drawn from the same distribution and the alternative hypothesis, $H_{1}$, is that the samples belong to two different distributions. This test is particularly valuable here as it helps us statistically assess the significance of the difference observed in the degree of inflation, $R_{\rm p, t}/R_{\rm p, 0}$ between the aligned and misaligned planet populations.

The right panel of Figure \ref{fig:radius inflation} presents the cumulative distribution functions (CDFs) of the degree of inflation for the aligned and misaligned planet populations, depicted by the red and blue curves, respectively. The \textit{K-S test} returns a p-value of approximately $0.099$, which allows us to reject the null hypothesis at the $\sim90\%$ confidence level (p-value $\le 0.1$). Consequently, we conclude that the aligned and misaligned planet populations are drawn from different distributions, with misaligned planets exhibiting a tendency for greater radius inflation.

\subsubsection{T-Test}
The \textit{t-test} is a parametric test designed to assess whether the means of two populations are equal under the assumption that both samples follow a normal distribution and have equal variances. This test is particularly effective for small samples, like ours, which consists of approximately a dozen planets in each sub-population. Even though small samples may deviate from perfect normality, the t-distribution can still provide a reasonable approximation, making the \textit{t-test} an appropriate choice for our analysis.

Performing a \textit{t-test} on the difference in the degree of inflation between the aligned and misaligned planet populations yields a p-value of $0.022$. Given this result, we reject the null hypothesis at the 97\% confidence level (p-value $\le 0.03$), indicating that the means of the two populations differ significantly. To ensure the robustness of these findings, we also perform \textit{Welch's t-test}, a variation of the \textit{t-test} that does not assume equal variances between the populations. \textit{Welch's t-test} produces a p-value of $0.024$, further reinforcing our conclusion. Before concluding, we try one more test.

\subsubsection{Mann and Whitney U-test}
The \textit{Mann-Whitney U-test} is a non-parametric test used to evaluate whether there is a significant difference between two populations by comparing their samples. Unlike the parametric \textit{t-test}, it does not require the data to follow a normal distribution, making it more robust in cases where the data is skewed or contains outliers. This makes the \textit{Mann-Whitney U-test} particularly well-suited for small sample sizes where traditional parametric tests may not perform reliably. 

Although the \textit{K-S test} is also non-parametric, these tests have key differences in what they assess and how they work. The \textit{Mann-Whitney U test} compares the central tendencies of two populations, ranking all values in both populations and comparing their ranks to determine whether one population generally has higher or lower values than the other. However, it does not assess the overall shape of the distributions. In contrast, the \textit{K-S test} compares the distributions as a whole, assessing whether the samples come from the same distribution by comparing the CDFs of the two populations. The \textit{K-S test} is sensitive to differences in both location and shape across the entire distribution, rather than just central tendency.

The \textit{Mann-Whitney U-test} applied to the aligned and misaligned planet populations yields a p-value of $0.046$, allowing us to reject the null hypothesis with $95\%$ confidence. These statistical results, consistent with the findings from our earlier tests, lead us to conclude that the aligned and misaligned planet populations are distinct, with misaligned planets experiencing significantly greater radius inflation due to tidal heating. A summary of the results from all of the statistical tests is given in Table \ref{tab:satistical results}. 

\begin{table}[] 
    \centering
    \begin{tabular}{ccccc}
    \toprule
        \textbf{Statistical Test} & \textbf{Test Stat.} & \textbf{p-value} & $\mathbf{H_{0}}$ & $\mathbf{H_1}$\\
        \textit{K-S Test} & 0.500 & 0.099 & \ding{55} & \checkmark \\
        \textit{Simple t-test} & 2.469 & 0.022 & \ding{55} & \checkmark \\
        \textit{Welch's t-test} & 2.469 & 0.024 & \ding{55} & \checkmark \\
        \textit{Mann-Whitney U-test} & 107 & 0.046 & \ding{55} & \checkmark \\
        \hline \hline 
    \end{tabular}
    \caption{Summary of results from statistical tests, all rejecting the null hypothesis $H_0$ with at least $90\%$ confidence, revealing that misaligned planets are generally more inflated than aligned ones.}
    \label{tab:satistical results}
\end{table}

A point to note is that, based on the classification criteria defined in Section~2, K2-261 b is placed in the aligned population since its measured sky-projected obliquity is $ \lambda = 32^\circ < 40^\circ$. However, this value is significantly larger than that of other aligned planets in our sample, and the true 3D obliquity of this system remains unknown. Although it is possible that the planet is truly aligned, it is also plausible that its true obliquity might place it among the misaligned population. To test the impact of this ambiguity, we re-ran all statistical tests excluding K2-261 b, and found that the confidence level for distinguishing the two populations increases to $\sim 97\%$.

Furthermore, for planets in our sample without measured eccentricities, we assumed a small nominal value of $e = 0.05 \pm 0.02$, as described in Section \ref{sec: mcmc}. This assumption allowed us to construct a sufficiently large sample for meaningful statistical analysis under the assumption of modest eccentricities. To assess sensitivity to this assumption, we repeat all statistical tests on a more conservative subsample that only includes planets with measured eccentricities (this subsample includes K2-261 b). We find that the main trend persists with a confidence level of at least 93\%, indicating that our conclusions are robust to this assumption.

\section{Case Study: WASP 107 \MakeLowercase{b}} \label{sec:wasp107}
One of the polar planets in our sample, WASP-107~b, has recently garnered significant attention. WASP-107~b is a close-in ($P_{\rm orb} = 5.72$ days), Neptune-mass ($M_p = 30.5 \ M_\oplus$) exoplanet with a radius comparable to that of Jupiter ($\sim 10 \ R_\oplus$), indicating an extremely low density ($\sim 0.12$ g/cm$^3$) and a substantial gaseous envelope \citep{2017A&A...604A.110A,
2017AJ....153..205D,
2021AJ....161...70P}. Observations indicate a small eccentricity of $e_0 = 0.06\pm0.04$ \citep{2021AJ....161...70P}. Although the orbit may be consistent with a circular configuration within $2\sigma$, we adopt $e_0 = 0.06$ for our analysis here and assume that the planet has not yet fully circularized. 

Our results identify WASP-107 b as the most inflated of all the misaligned planets in our sample with $R_{\rm p,t}/R_{\rm p,0} \approx 1.9$. A high heat flux from tidal heating provides a compelling explanation for the planet's inflated radius \citep{Millholland2020}. This interpretation has been recently supported by observations from the James Webb Space Telescope, which revealed an unexpectedly high internal temperature for WASP-107~b, consistent with significant tidal heating \citep{2024Natur.630..831S, 2024Natur.630..836W}. For instance, \cite{2024Natur.630..836W} measured an internal temperature $T_{\mathrm{int}}\gtrsim345$ K, yielding a tidal quality factor $Q \lesssim 10^{3.8}$. Given its intriguing characteristics, here we delve deeper into studying the dynamical evolution of this planet and couple it with an analysis of its physical properties. We aim to assess the timescale and efficiency of its tidal evolution.

The WASP-107 system also contains a distant companion, WASP-107 c, which could haven driven high-eccentricity migration of WASP-107~b. \citet{2024ApJ...972..159Y} modeled this process by coupling octupole-order von Zeipel-Lidov-Kozai (ZLK) oscillations with various short-range forces and demonstrated that high-eccentricity migration can indeed explain the peculiar orbital architecture and polar orbit of WASP-107~b. In our simplified model, we will not study these three-body interactions and solely focus on the dynamical evolution of WASP-107~b after the ZLK oscillations have been suppressed. At the end of the ZLK process, the planet would be left with a highly eccentric orbit that gradually circularizes due to tides. Our model captures this phase of orbital evolution.
\subsection{Tidal Orbital Evolution} \label{subsec:orb_dyn}
We model the orbital dynamics of WASP 107 b using the framework from \citet{2010A&A...516A..64L}, which follows a traditional viscous approach of equilibrium tide theory. The semi-major axis evolution is given by 
\begin{align} \label{da_dt}
       \frac{1}{a}\frac{da}{dt} = \frac{4a}{GM_\star M_p}\Bigg\{ K\left [N(e)x_p\frac{\omega_p}{n} - N_a(e) \right] 
    \nonumber\\ 
    K_\star \left [ N(e) x_\star \frac{\omega_\star}{n} - N_a(e)\right] \Bigg\}.
\end{align}
Here $N_a(e)$, $N(e)$ and $K$ are as defined in equations \ref{n_a}, \ref{n(e)} \& \ref{K}, $\omega_p$ is the planet’s rotation rate, and $x_p = \cos\epsilon$, where $\epsilon$ is the planetary obliquity. For simplicity and consistency with previous results, we assume $\epsilon = 0$ in our analysis. We also assume the spin rate is at equilibrium, $\omega_p = \omega_{\mathrm{eq}}$ (equation \ref{eq: omega_eq}).
All parameters with a ``$\star$'' subscript correspond to the stellar counterparts of the respective planetary parameters.
Similarly the secular evolution of eccentricity is given as
\begin{align} \label{de_dt}
       \frac{1}{e}\frac{de}{dt} = \frac{11a}{GM_\star M_p}\Bigg\{ K\left [\Omega_e(e)x_p\frac{\omega_p}{n} - \frac{18}{11}N_e(e) \right] 
    \nonumber \\ 
    K_\star \left [ \Omega_e(e) x_\star \frac{\omega_\star}{n} - \frac
    {18}{11}N_e(e)\right] \Bigg\},
\end{align}
where $\Omega_e(e)$, and $N_e(e)$ are defined as 
\begin{equation}
    \Omega_e(e) = \frac{1 + \frac{3}{2}e^2 + \frac{1}{8}e^4}{(1 - e^2)^5}
\end{equation}
\begin{equation} 
    N_e(e) = \frac{1 + \frac{15}{4}e^2 + \frac{15}{8}e^4 + \frac{5}{64}e^6}{(1-e^2)^{13/2}}.
\end{equation}
The terms proportional to $K$ and $K_{\rm \star}$ represent the effects of tides raised by the star on the planet and by the planet on the star, respectively. Since $K_\star/K \lesssim 10^{-4}$, the tidal effects exerted by the planet on the star are negligible compared to the tides exerted by the star on the planet. Therefore, we ignore the terms proportional to $K_{\rm \star}$ in both equations \ref{da_dt} \& \ref{de_dt}. This makes it such that we cannot evolve the stellar obliquity, but recent work by \cite{2024ApJ...974..304L} found that WASP-107~b's polar orbit is long-term stable.

As mentioned above, we investigate the orbital evolution of WASP-107~b in a phase where it is undergoing isolated tidal evolution, subsequent to a possible history of ZLK oscillations. At the start of this phase, the planet is envisioned to be stranded with a highly eccentric orbit, which would then evolve under tidal forces. We explore this by solving the two coupled differential equations described earlier (equations \ref{da_dt} \& \ref{de_dt}). We consider a grid of initial conditions, varying $e$ and $Q'$. The eccentricity is varied within the range, $e \in [0.98, 0.3]$, while $Q'$ spans the range 
$Q'$ $\in [10^3, 10^7]$. The semi-major axis corresponds to a given $e$ using the conservation of angular momentum with the planet's present-day conditions: $L(e, a) = \sqrt{GM_\star M_p^2 a (1 - e^2)} = L(e_0, a_0)$ where $e_0 = 0.06$, and $a_0 = 0.055$ AU, are the present-day eccentricity and semi-major axis, such that 
\begin{equation}
\label{a_grid}
    a = \frac{a_0(1 - e_0^2)}{(1 - e^2)}.
\end{equation}

Our first goal is to investigate how long it has been since the ZLK oscillations were suppressed. We therefore solve the coupled differential equations across the grid of initial conditions specified above, over a timespan of 5 Gyr. For each initial condition, we identify the point along the evolutionary trajectory that is closest to the present-day orbital orientation of the system. Specifically, we find the point, ($a_m, e_m$) along the trajectory, that minimizes the function, $f(a,e) = \sqrt{(a - a_0)^2 + (e - e_0)^2}$. If $f(a_m, e_m) \le 10^{-3}$, we consider the system to have converged to the present-day conditions and record the time, $t$, at which this convergence occurs. 

\begin{figure}
    \centering
    \includegraphics[width = 1.02\linewidth]{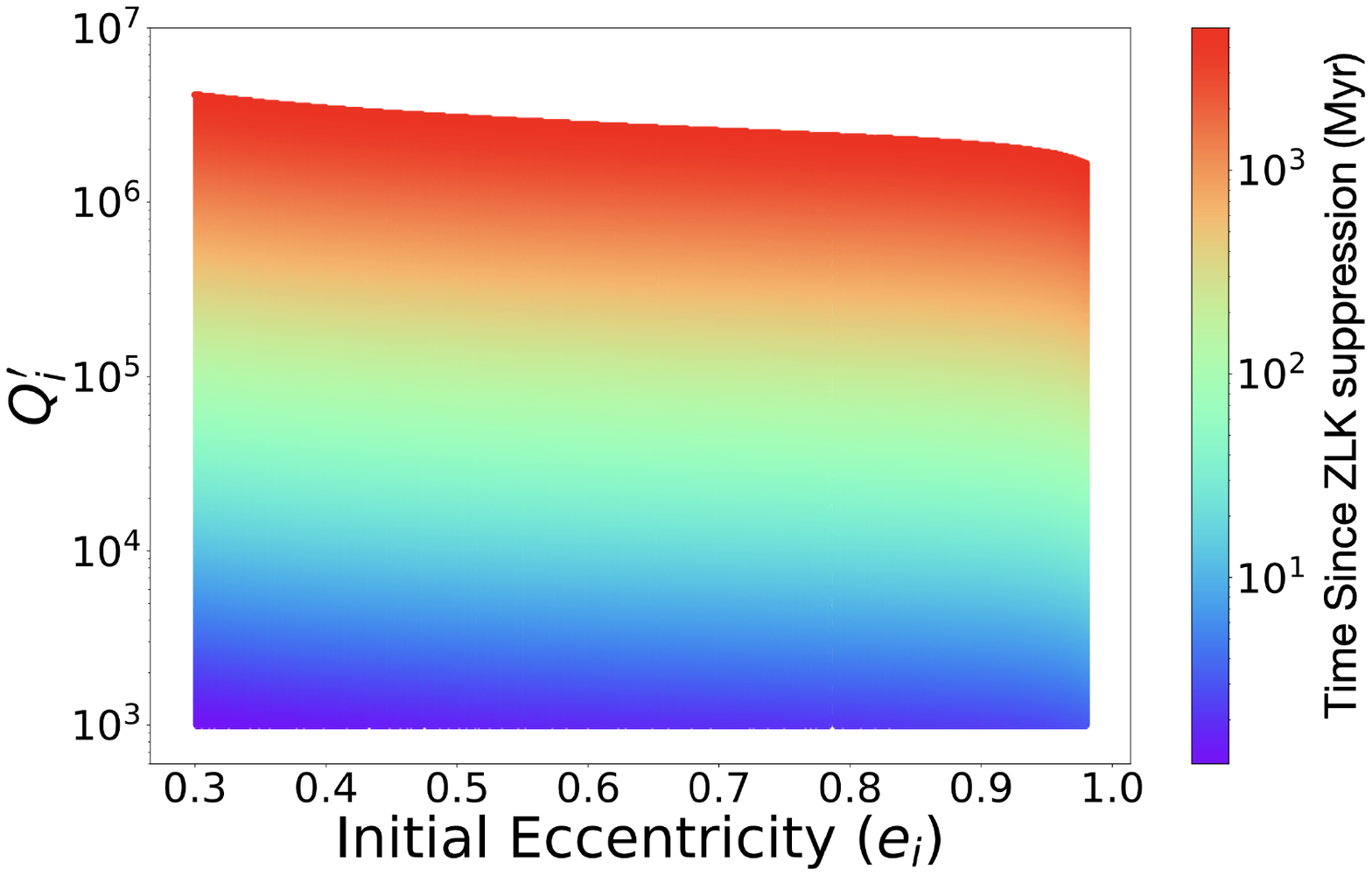}
    \caption{Results of tidal orbital simulations of WASP-107~b. The colorbar indicates the time since the suppression of ZLK oscillations, starting from an initial state ($e_i, Q'_i$) after the suppression.}
    \label{fid:initial_grid}
\end{figure}

Figure \ref{fid:initial_grid} shows the results of these calculations, indicating the time required for the planet to reach its present day conditions, starting from an initial state $(e_i, Q'_i)$ after the suppression of ZLK oscillations. We observe that as $e_i$ increases, there is a slight increase in $t$ values. According to equation \ref{a_grid}, a higher $e_i$ should result in a larger $a_i$. Since tidal heating is less effective at larger semi-major axes, the circularization process slows, leading to longer evolution times. We also observe a smooth increase in $t$ as $Q'_i$ increases from $10^3$ to $10^7$. For $Q'_i \approx 10^3$, $t$ is only a few Myr due to rapid tidal circularization. Around $Q'_i \approx 10^6$, $t \gtrsim 1$ Gyr, and for very high $Q'_i$ (around $10^7$), none of the initial conditions converge to the present-day values within the simulation timeframe of 5 Gyr. For such high $Q'$ values, tidal dissipation is weak enough that the planet would require more than $5$ Gyr to reach its current state. 

Given the estimated age of this system, $3.4 \pm 0.7$ Gyr \citep{2023A&A...669A..63B}, we rule out excessively high $Q'$ values ($\gtrsim 10^7$). Similarly, low $Q'$ values ($Q' \lesssim 10^4$) are disfavored because they suggest a rapid circularization time on the order of Myr, inconsistent with the measured marginal non-zero eccentricity. Moreover, according to \citet{2024ApJ...972..159Y}, the timescale of high-eccentricity migration via ZLK cycles requires approximately 100 Myr. If $Q' \lesssim 10^4$, the system would have circularized well before the present day, which is inconsistent with observations. Thus, our results favor $Q'\approx10^4 - 10^6$, which aligns well with the findings of \citet{2024ApJ...974..304L}, who explored this planet with an independent analysis. It also matches well with the constraints imposed by recent JWST observations \citep{2024Natur.630..831S, 2024Natur.630..836W}, discussed in more detail in Section \ref{sec-results}.  
\subsection{Simultaneous Radius and Orbital Evolution}
While the dynamical simulations we just presented are a useful start, they do not provide a complete picture. As discussed in previous sections, tidal heating significantly impacts a planet's physical properties, especially through radius inflation. The orbital and physical evolution is actually coupled, since the extent of radius inflation is a function of the semi-major axis and eccentricity, and simultaneously $da/dt$ and $de/dt$ are a function of the radius. While it is a simplifying approximation to keep the planet's radius fixed during the orbital evolution, a more accurate calculation would have $R_p$ evolve simultaneously with $e$ and $a$. For example, recent studies  \citep[e.g.][]{2025ApJ...979..218L} accounted for radius evolution in their models for studying dynamics of another polar planet, HAT-P-11 b.

\subsubsection{Instantaneous radius evolution}
To incorporate this radius inflation into the dynamical model, we first assume a simplified picture where the planet's radius instantaneously adjusts to its equilibrium radius, $R_{\rm eq}$, which is the radius predicted by our models that include tidal heating. We calculate $R_{\rm eq}$ by interpolating MESA simulations for models accounting for tidal heating, as discussed in Section \ref{sec: mcmc}. Recalling that the model including tides depends on four parameters summarized in Table \ref{tab:parameters}, $R_{\rm eq}$ also depends on these four parameters: $M_p$; $F/F_\oplus$, which is a function of $a$; $f_{\rm env, t}$, which we estimate for a given $Q'$ using the method described in Section \ref{sec:f_env}; and $\Gamma$, which depends on $Q'$ as well as $a$ and $e$. 

For this analysis, we fix $Q' = 10^6$ based on the constraints described in Section \ref{subsec:orb_dyn}. Using the method detailed in Section \ref{sec:f_env}, we determine the value of $f_{\rm env, t}$ corresponding to this $Q'$ to be $\geq 50\%$. However, since our planetary evolution model is defined only within a range of $f_{\rm env}$ values with a maximum limit of $50\%$, we fix $f_{\rm env, t} = 0.49$ to ensure compatibility with the model. Using the interpolated MESA simulations, we determine the $R_{\rm eq}$ corresponding to the evolving values of $a$ and $e$, while fixing $Q'$. In this simplified model, $R_p$ is assumed to instantaneously adapt to $R_{\rm eq}$ at each step of solving the coupled differential equations \ref{da_dt} and \ref{de_dt}. We call this approach the ``\textit{instantaneous radius evolution}''.

\subsubsection{Radius evolution with time lag}
However, we know that the inflation or deflation process is not instantaneous and it requires time to achieve equilibrium. This delay leads to a time lag in the radius evolution. To address this limitation in our simplified model, we create a third approach by introducing a differential equation that governs the radius evolution \citep{2024ApJ...972..159Y},
\begin{equation} \label{dr_dt}
    \frac{dR_p}{dt} = \frac{R_{\rm eq}(t) - R_p}{\tau}
\end{equation}
\begin{equation}
    \tau = \begin{cases}
 \tau_{\rm inf}, & R_p < R_{\rm eq}(t)\\
\tau_{\rm def}, & R_p > R_{\rm eq}(t)
\end{cases}
\end{equation}
where $\tau_{\rm inf}$, and $\tau_{\rm def}$ are the inflation and the deflation timescales, respectively. This equation was utilized by \citet{2024ApJ...972..159Y} in their study of WASP-107 b's high-eccentricity migration via ZLK oscillations. However, they derived $R_{\rm eq}$ using a parameterized radius prescription provided by \citet{2021ApJ...909L..16T}. In contrast, our analysis determines $R_{\rm eq}$ directly from the interpolation to the MESA simulations. 

\citet{2021ApJ...909L..16T} found that the cooling timescales (for hot Jupiters) are significantly slower, whereas their reinflation occurs much more rapidly. Following their findings, we adopt the same deflation timescale, $\tau_{\rm def} = 0.5$ Gyr, as used in their analysis. Additionally, for our simulations, we assume an inflation timescale, $\tau_{\rm inf} = 5.8$ Myr, consistent with the value utilized in specific simulations conducted by \citet{2024ApJ...972..159Y}. Solving the three coupled differential equations, \ref{da_dt}, \ref{de_dt} and \ref{dr_dt} simultaneously allows us to model the planet's radius and orbital evolution concurrently. We refer to this method as the ``\textit{radius evolution with time lag}''. 

\begin{figure} 
    \centering
    \includegraphics[width=\linewidth]{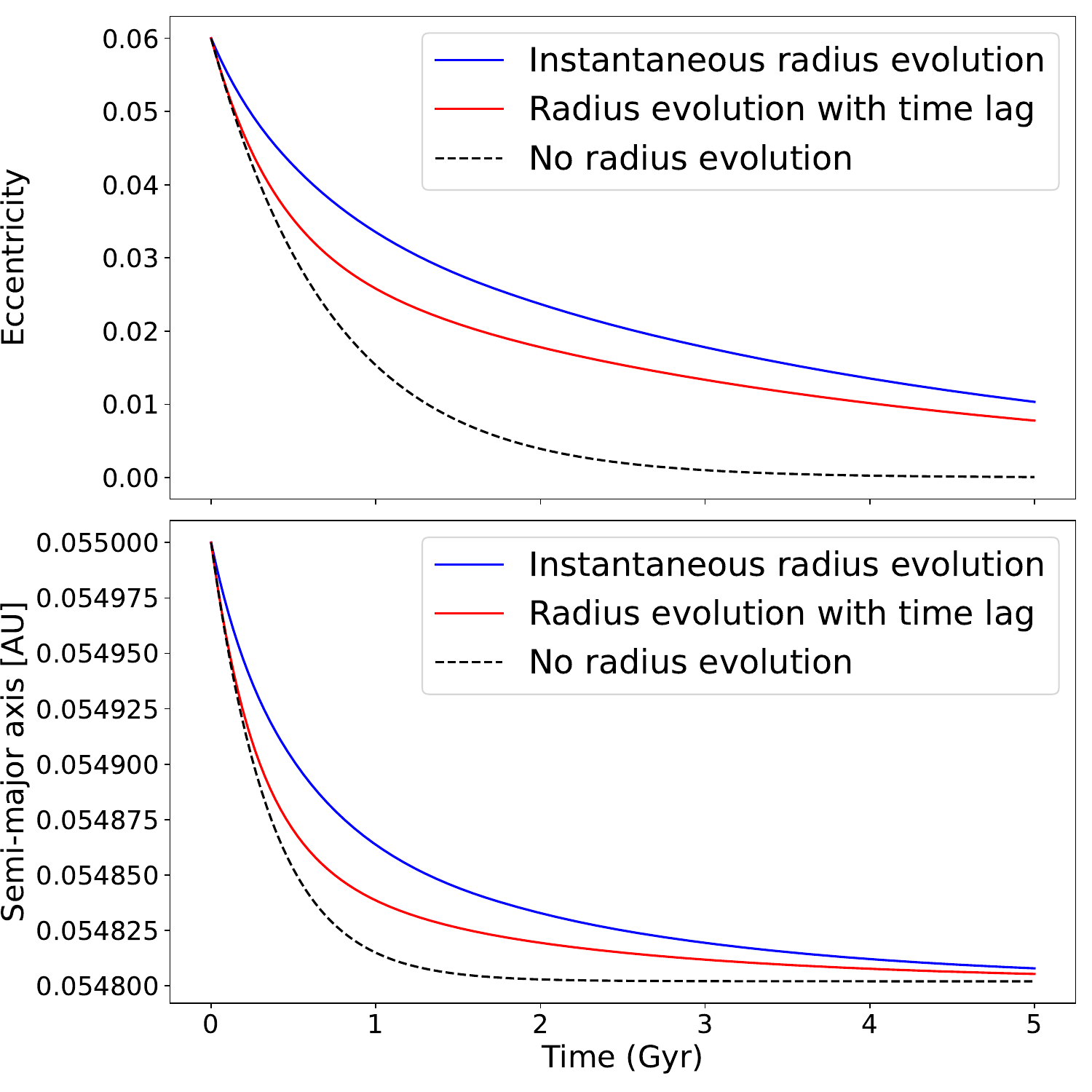}
    \caption{Tidal orbital evolution of WASP-107~b using the \textit{no radius evolution} (dashed black curve), \textit{instantaneous radius evolution} (blue curve), and \textit{radius evolution with time lag} (red curve) approaches for $Q' = 10^6$. We show the evolution of the eccentricity (top) and semi-major axis (bottom) over a period of 5 Gyr starting from the present-day state of the planet ($e_0, a_0, R_{\rm p,0}$).}
    \label{fig_orb_dyn}
\end{figure}

\subsubsection{Results} \label{sec-results}
We model the orbital dynamics of WASP-107~b using the three approaches outlined above: no radius evolution, instantaneous radius evolution, and radius evolution with time lag. Starting from the observed present-day parameters ($a_0$, $e_0$, $R_{\rm p,0}$), we evolve the system forward in time over a period of 5 Gyr. The results of these simulations are presented in Figure \ref{fig_orb_dyn}. They demonstrate that, with $Q' \approx 10^6$ (the upper plausible limit of $Q'$ derived in Section \ref{subsec:orb_dyn}), radius evolution introduces a noticeable impact on the orbital dynamics. 

The system evolves more slowly, retaining a small eccentricity even after 5 Gyr. In contrast, the model without radius evolution predicts a fully circular orbit after $\sim 3$ Gyr. As expected, the radius evolution with time lag method produces results that fall between those of the no radius evolution and instantaneous radius evolution models. 

The distinct differences in orbital evolution when radius changes are accounted for highlight the importance of incorporating both orbital and physical effects to achieve a more accurate dynamical history. These findings also align well with observations of WASP 107 b's slightly eccentric orbit, and they reinforce the plausibility of a $Q'$ value around $10^4-10^6$. Furthermore, \citet{2024Natur.630..831S}, used JWST NIRSpec data and reported a core mass of $M_{\rm core}=11.5^{+3}_{-3.6} M_\oplus$. This implies an envelope mass of approximately $M_{\rm env} = M_p - M_{\rm core} = 19 \ M_\oplus$, resulting in $f_{\rm env} > 50\%$. According to our planet evolution model incorporating tides, such a high $f_{\rm env}$ corresponds to $Q' \gtrsim 10^4$ (refer to Figure \ref{fig:f_env sim}, top panel, fourth plot corresponding to WASP-107~b). Similarly, \cite{2024Natur.630..836W} inferred $Q \sim 10^{3.8}$ (which would yield $Q' = 3Q/2k_2$ larger than this). Both results are consistent with our predictions of $Q'$ from our dynamical analysis. 

While this paper was in review, we also became aware of recent work by \cite{2025ApJ...985...87B} that argued that ohmic dissipation offers a better explanation for WASP-107 b's properties than tidal heating. As mentioned earlier, we did not consider ohmic heating in this work, but we suggest that future studies of close-in, inflated Neptunes/sub-Saturns should evaluate multiple sources of internal heating when attempting to explain the nature of these planets.

\section{Conclusions} \label{sec:conclusions}
Motivated initially by the peculiar architectures of polar-orbiting planets, we investigated a sample of 24 planets to search for evidence of a relationship between orbital misalignments and tidally-induced radius inflation. We showed that misaligned planets experience greater average tidal dissipation and, consequently, more significant radius inflation compared to aligned planets. The key analyses and findings of this study are summarized as follows:

\begin{enumerate}
    \item We compiled a sample of 12 misaligned and 12 aligned planets, detailed in Table \ref{tab:planet sample}. We characterized each planet with two planetary structure models: one including tidal heating and the other excluding it. We calculated the envelope mass fraction estimates: $f_{\rm env, t}$ from the model with tides, and $f_{\rm env, 0}$ from the model without tides (reported in Table \ref{tab:radius inflation} for $Q' = 10^4$). For all planets, $f_{\rm env, t} < f_{\rm env, 0}$ since a tidally inflated planet model requires a smaller envelope mass to match the observed radius. The average difference between the envelope mass fractions is $\langle f_{\rm env, 0} - f_{\rm env, t} \rangle_a \approx 5\%$ for the aligned planets and $\mathbf{\langle f_{\rm env, 0} - f_{\rm env, t} \rangle_m \approx 12\%}$ for the misaligned planets. This disparity suggests a more important role of tidal inflation for the misaligned planets.
    
    \item We further quantified this with the degree of inflation, $R_{\rm p,t}/R_{\rm p,0}$ (reported in Table \ref{tab:radius inflation} for $Q' = 10^4$). This indicates the extent that an observed planet's radius has been affected by tides (i.e. ${R_{\rm p,t}/R_{\rm p,0}=1}$ indicates no inflation). The average degree of inflation is $\langle R_{\rm p, t}/R_{\rm p, 0} \rangle_a \approx 1.1$ for the aligned planets and $\langle R_{\rm p, t}/R_{\rm p, 0} \rangle_m \approx 1.4$ for the misaligned planets, indicating more substantial inflation for the misaligned planets.
    
    \item We used the distribution of computed $R_{\rm p, t}/R_{\rm p, 0}$ values to evaluate the statistical significance of the difference in the degree of inflation between the aligned and misaligned planets. Four different statistical tests (refer to Table \ref{tab:satistical results} for a summary) confirm with at least $90\%$ confidence that the misaligned planets are more inflated due to tidal effects than the aligned planets, supporting our conclusion that the two populations exhibit distinct tidal behaviors.

    \item Our finding that misaligned planets tend to be more inflated than aligned ones suggests that these planets may have experienced dynamically violent processes that drove them onto eccentric orbits, where strong tidal dissipation would naturally lead to increased radius inflation. In contrast, if the primary driver of misalignment was one where the protoplanetary disk itself became tilted relative to the stellar spin axis, then one would not expect a significant difference in tidal heating or inflation between the aligned and misaligned populations. In such scenarios, misaligned planets would not have been subjected to the high eccentricities that induce substantial tidal heating.
    
    \item We examined a case study of the most inflated planet in our sample, WASP-107~b. We modeled its coupled tidal orbital dynamics and radius evolution, treating the planet to be evolving independently, subsequent to a potential history of ZLK oscillations with its long-period companion that drove high-eccentricity migration \citep{2024ApJ...972..159Y}. For $Q'\lesssim 10^4$, the planet would have exited ZLK oscillations only within the last few Myr. This is challenging to reconcile with the system's estimated age ($3.4 \pm 0.7$ Gyr; \citealt{2023A&A...669A..63B}) and the findings of \citet{2024ApJ...972..159Y}, which suggested that ZLK oscillations would have taken place over $\sim100$ Myr. Conversely, $Q'$ cannot be excessively large either, as very high $Q'$ values ($Q' \gtrsim 10^6$) would result in tidal circularization occurring too slowly, requiring a timescale greater than the system's age to reach the present-day orbital state. Our results thus suggest that $Q' \approx 10^4 - 10^6$ for WASP-107~b, striking a balance between rapid tidal circularization and the system's present-day orbital architecture. However, ohmic dissipation (not explored in this work) is a compelling potential alternative to tidal heating in this planet \citep{2025ApJ...985...87B}.
    
\end{enumerate}

Our results provide evidence of a statistically significant correlation between orbital misalignment and enhanced tidal inflation. Still, future observations and an expanded dataset can strengthen this conclusion. Our findings also highlight the importance of incorporating radius evolution within orbital modeling to achieve a more accurate history of tidally inflated planets like WASP-107~b. There is a clear need for sophisticated models that better account for this coupled orbital-interior evolution. We leave the development of such models to future studies. These could then be applied to a larger sample of aligned and misaligned planets, enabling a more comprehensive understanding of how tidal evolution has played a role in sculpting the orbital and physical states of the two populations.

\section*{Acknowledgements}
We thank the anonymous reviewer for their helpful comments that improved this paper. This material is based upon work supported by the National Science Foundation under Grant No. 2306391. This research has made use of the NASA Exoplanet Archive, which is operated by the California Institute of Technology, under contract with the National Aeronautics and Space Administration under the Exoplanet Exploration Program.

\bibliographystyle{aasjournalv7}
\bibliography{main}

\end{document}